# Atomically resolved intrinsic superconducting gap in $(La,Pr)_3Ni_2O_7$ films

Xinxin Wang[1,†], Yaqi Chen[1,†], Cui Ding[1,2,†,*], Lizhi Xu[1,†], Jian-Jian Miao[1,2], Guangdi Zhou[1,2,*], Zhuoyu Chen[1,2], Yu-Jie Sun[1,2,*], Jin-Feng Jia[1,2,3,*] & Qi-Kun Xue[1,2,4,*]

[1]State Key Laboratory of Quantum Functional Materials, Department of Physics, Center for Advanced Light Source, Guangdong Basic Research Center of Excellence for Quantum Science, and College of Semiconductors (National Graduate College for Engineers), Southern University of Science and Technology, Shenzhen 518055, China

[2]Quantum Science Center of Guangdong-Hong Kong-Macao Greater Bay Area, Shenzhen, 518045, China

[3]State Key Laboratory of Micronano Engineering Science, Tsung-Dao Lee Institute and School of Physics and Astronomy, Key Laboratory of Artificial Structures and Quantum Control (Ministry of Education), Shanghai Jiao Tong University, Shanghai, 200240 China.

[4]Department of Physics, Tsinghua University, Beijing, 100084, China

[†]These authors contributed equally: Xinxin Wang, Yaqi Chen, Cui Ding, Lizhi Xu.

*e-mail: dingcui@quantumsc.cn; zhouguangdi@quantumsc.cn; sunyj@sustech.edu.cn; jiajf@sustech.edu.cn; xueqk@sustech.edu.cn.

## Abstract

Ruddlesden–Popper bilayer nickelates provide an emerging platform for studying high-temperature superconductivity, yet the superconducting pairing symmetry remains under debate. Here, we use atomic-resolution scanning tunnelling microscopy and spectroscopy to investigate superconducting 1.5-unit-cell $(La,Pr)_3Ni_2O_7$ films grown on $SrLaAlO_4$. A cryogenic ultrahigh-vacuum (UHV) sample transfer preserves an ordered $\sqrt{2} \times \sqrt{2}$ surface and yields reproducible U-shaped spectra with two gap scales of ~14 and ~20 meV and extended flat zero-conductance bottoms. By contrast, samples exposed for a longer time in UHV without cooling during transfer show V-shaped spectra despite retaining the surface reconstruction and a transport superconducting transition onset above 40 K. Wide-energy-range spectra indicate that oxygen loss can mix density-wave-related spectral weight. Our measurements provide an atomic-scale observation of the intrinsic nodeless superconducting gap in bilayer nickelate ultrathin films.

## Introduction

Nickelates have rapidly emerged as a central platform for studying high-temperature superconductivity in addition to cuprates and iron-based superconductors. Superconductivity in nickelates was first discovered in square-planar infinite-layer films and was then extended to multi-layer systems[1–4]. In the Ruddlesden–Popper (RP) phases, superconductivity was first reported in pressurized bilayer single crystals, and later realized in epitaxial bilayer nickelate thin films at ambient pressure[5–13]. The RP family now spans pressurized trilayer single crystals, hybrid RP single crystals and thin films, and other chemically substituted or structurally engineered nickelates[14–18]. This broader materials landscape emphasizes that superconductivity in nickelates depends sensitively on layer number, structural polymorphism, strain, and oxygen content[19,20].

The central experimental issue for RP nickelate superconductors is the origin of superconductivity. Resonant inelastic X-ray scattering (RIXS), X-ray absorption spectroscopy (XAS) and other measurements show strong Ni–O hybridization, magnetic excitations and ligand-hole states that are sensitive to oxygen stoichiometry[21–25]. Momentum-resolved angle-resolved photoemission spectroscopy (ARPES) and related electronic-structure studies on superconducting bilayer thin films have resolved orbital-dependent correlations, reconstructed low-energy bands, Fermi-surface sheets, nodeless gap behaviour and gap-related spectral changes near the Fermi level[26–34]. In particular, the ARPES/RIXS study across the superconductor–insulator transition in bilayer films found that oxygen loss reorganizes unoccupied orbital character and induces density-wave orders[30-32]. These observations indicate that oxygen control is not merely a growth detail, but part of the low-energy electronic problem that must be controlled before the superconducting gap symmetry can be inferred.

Here, we combine atomic-layer-controlled growth, cryogenic ultrahigh-vacuum (UHV) sample transfer and scanning tunnelling microscopy and spectroscopy (STM/STS) to investigate microscopic superconducting properties of 1.5-unit-cell (equivalent to 3 bilayers) $(La,Pr)_3Ni_2O_7$ ultrathin films. We show that minimizing the sample transfer time in UHV without cooling (between removal from the cryogenic UHV suitcase and arrival in the STM low-temperature sample stage) produces homogeneous U-shaped double-gap spectra with extended flat zero-conductance bottoms, whereas longer-time transfer yields V-shaped spectra with a broad dip characterized by density-wave energy scales. This transfer-time comparison identifies oxygen control as a prerequisite for accessing the intrinsic superconducting gap and provides a practical criterion for interpreting local spectra in bilayer nickelate films.

## Results

### Atomically resolved superconducting bilayer nickelate films

Using the gigantic-oxidative atomic-layer-by-layer epitaxy (GAE) method[35], we synthesized 1.5-unit-cell $(La,Pr)_3Ni_2O_7$ ultrathin films on treated (001)-oriented $SrLaAlO_4$ substrates. The nominal La:Pr ratio is 1.35:1.65 for all samples, and the top-layer sequence follows (La,Pr)O-$NiO_2$-(La,Pr)O-$NiO_2$, leaving the as-grown surface terminated by a Ni–O plane, as shown in Fig. 1a. Reflection high-energy electron diffraction during growth shows sharp diffraction spots and persistent oscillations with negligible decay, indicating layer-by-layer growth and high crystallinity (Extended Data Fig. 1a). X-ray diffraction confirms phase purity and coherent epitaxy (Extended Data Fig. 1b). A key step is the rapid cooling of as-grown films below 150 K followed by transfer in an ultrahigh-vacuum cryogenic suitcase.

Large-scale topographs show atomically flat terraces separated by $SrLaAlO_4$ unit-cell-scale steps (Fig. 1c and Extended Data Fig. 1c), indicating that the film remains morphologically coherent after transfer. On each terrace, atomic-resolution images reveal a clear $\sqrt{2} \times \sqrt{2}$ reconstruction relative to the underlying Ni–O lattice (Fig. 1d, e). This reconstruction is reproducible on rapidly transferred samples and provides an internal marker that the surface termination is well ordered. Transport measured on films prepared under the same growth conditions shows an onset superconducting transition near 53 K (Fig. 1f), consistent with ambient-pressure superconductivity reported for compressively strained bilayer nickelate films[8–12]. These structural and transport results establish the starting point for local spectroscopy, but they do not by themselves determine whether the tunnelling gap is intrinsic; that question requires a direct comparison of spectra with different oxygen histories.

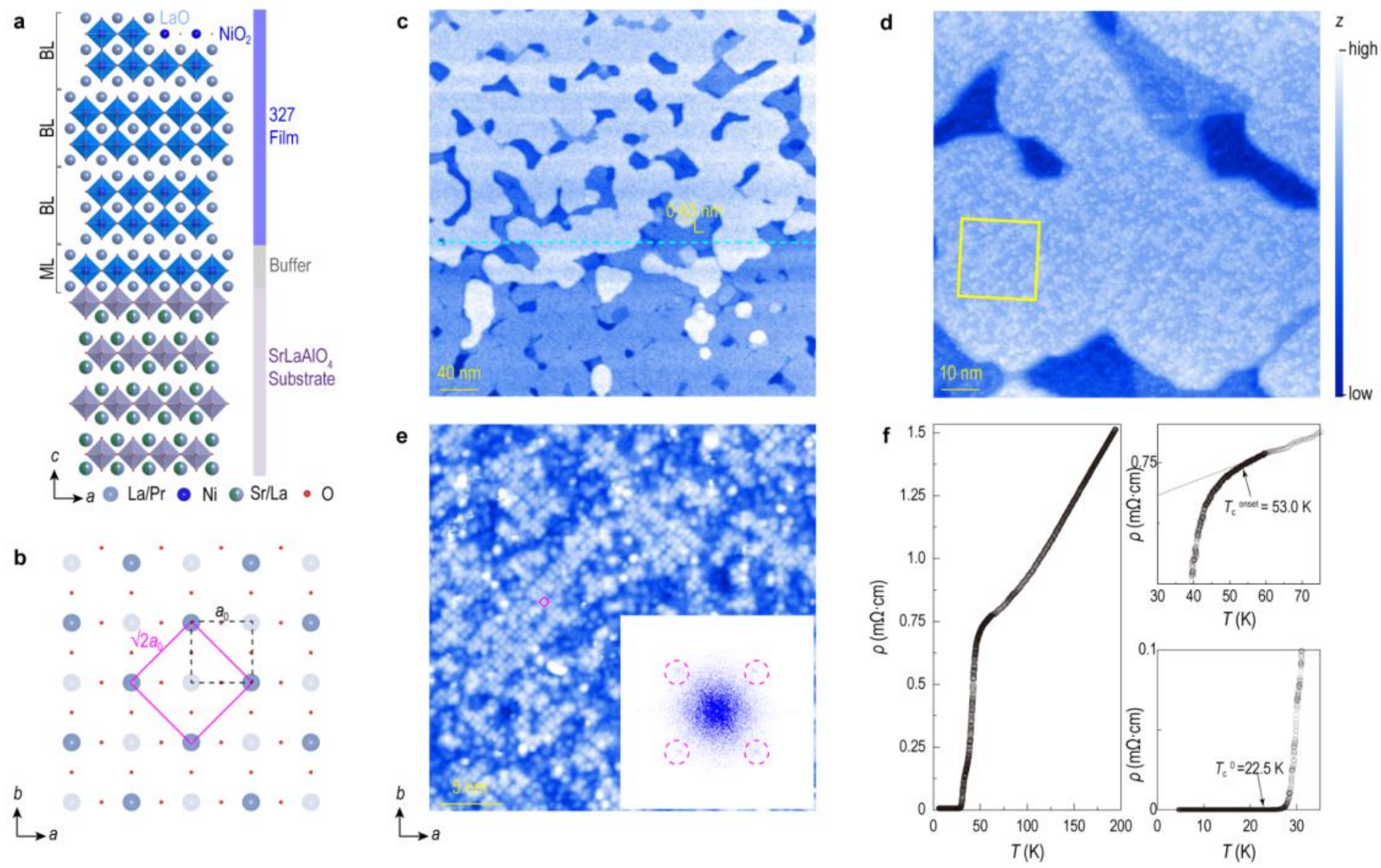

**Fig. 1 | Atomically resolved superconducting 1.5-unit-cell $(La,Pr)_3Ni_2O_7$ film grown on $SrLaAlO_4$.** **a**, Schematic crystal structure of the $(La,Pr)_3Ni_2O_7$ bilayer. **b**, Schematic illustration of the √2 × √2 surface reconstruction. The black dashed square marks the smallest lattice period and the purple solid square marks the reconstructed surface unit. **c**, Large-scale STM image (set point: $V_s$ = 2 V, $I_s$ = 50 pA), showing flat terraces and distinct steps. **d**, Zoomed-in STM image (set point: $V_s$ = 2 V, $I_s$ = 50 pA). **e**, Atomic-resolution STM image (set point: $V_s$ = 150 mV, $I_s$ = 10 pA). The inset shows the corresponding fast Fourier transform, with purple dashed circles marking the reconstructed Bragg points corresponding to a real-space periodicity of 0.55 nm. **f**, Temperature-dependent resistivity curve for the same growth batch, showing an onset superconducting transition near 53 K.

## U-shaped STS reveals nodeless double-gap superconductivity

We next probe the local electronic structure by spatially resolved STS on the atomically ordered terraces. Line-cut spectra acquired across an atomically resolved region (lower panel of Fig. 2a) show fully opened, U-shaped superconducting gaps with strongly suppressed zero-bias conductance (Fig. 2b). Two sets of coherence peaks are resolved within the same spectrum, corresponding to characteristic energy scales of about 12 and 20 meV. Spectra taken in a second area exhibit a similar U-shaped superconducting gap with inner and outer gap energies of about 14 and 20 meV (Fig. 2d). These results are consistent with recent ARPES and leading-edge spectroscopic evidence for superconducting bilayer nickelate films[27–30,33,34]. The gap energies and line shapes remain spatially homogeneous over the measured line cuts. The nearly vanishing low-bias conductance is the central feature: it shows that, in rapidly transferred films, the local density of states is fully depleted at the Fermi level rather than forming a simple V-shaped suppression.

The double-gap structure is naturally expected in a multiband bilayer system in which different Fermi-surface sheets and orbital components contribute to tunnelling with unequal matrix elements. We do not use the two peak energies alone to assign individual bands; instead, we emphasize the robust phenomenology that both peak sets coexist within one fully opened spectrum. This observation distinguishes the superconducting U-shaped spectra from partial-gap or pseudogap-like spectra in which residual low-energy density of states remains large. It also motivates a quantitative comparison with nodeless two-gap fits in Fig. 3.

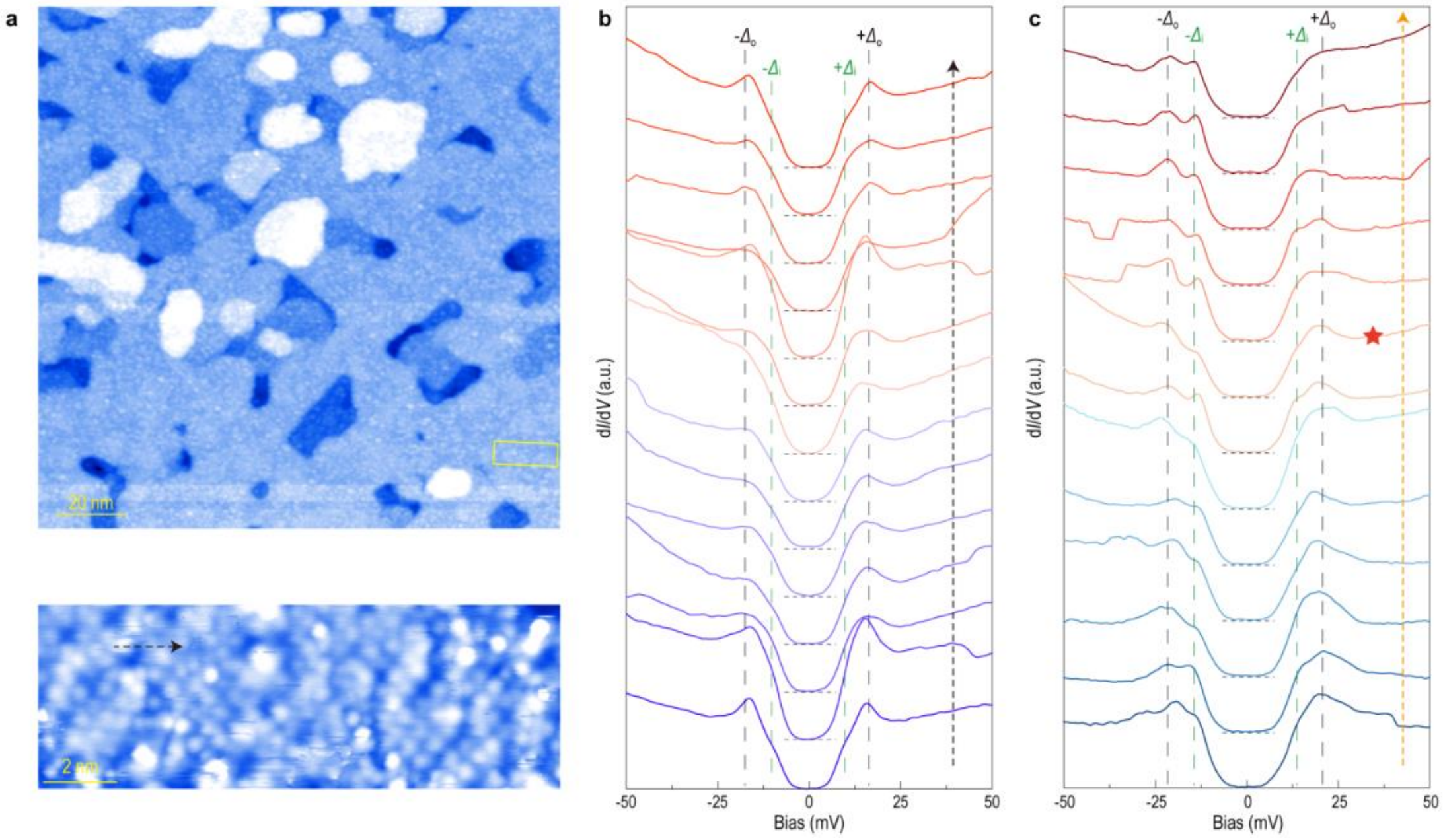

**Fig. 2 | Topographic and spectroscopic characterization of the superconducting double gap. a**, STM topographic image of a 200 x 200 nm$^2$ region (upper panel; set point: $V_s$ = 2 V, $I_s$ = 10 pA) and a zoomed-in image (lower panel; set point: $V_s$ = 900 mV, $I_s$ = 10 pA) of the area marked by the yellow rectangle. **b**, Tunnelling conductance spectra acquired along the black dashed arrow in **a** (set point: $V_s$ = 50 mV, $I_s$ = 300 pA, $V_{mod}$ = 1 mV). **c**, Tunnelling conductance spectra acquired along the yellow dashed arrow in **Extended Data Fig. 2d** (set point: $V_s$ = 50 mV, $I_s$ = 300 pA, $V_{mod}$ = 1 mV). Spectra are vertically shifted for clarity. Black and green vertical dashed lines indicate the outer and inner gap features, with gap energies of about 20 and 12–14 meV, respectively. **d**, Representative U-shaped spectrum from a second area, confirming the reproducibility of the nodeless double-gap line shape.

## Temperature evolution of U-shaped superconducting gap and gap fitting

Temperature-dependent spectra track the evolution of the superconducting state (Fig. 3a). With increasing temperature, the coherence peaks are gradually reduced and the zero-bias conductance increases, as expected for gap filling or closing. Nevertheless, the spectral line shape remains identifiable over the measured temperature range, indicating robust quasiparticle features in the strained bilayer thin films. To characterize the coherence peaks, we use the negative second derivative of symmetrized normalized d$I$/d$V$ curves and extract inner and outer gap energies, $\Delta_i$(T) and $\Delta_o$(T) (Extended Data Fig. 3). The temperature dependence can be described by BCS-like gap functions (Fig. 3b), giving characteristic temperatures of about 43 and 58 K, comparable to the transport onset temperature of 53 K. We further fit the normalized 4.2 K spectrum with a two-gap Dynes density of states[36,37]. A nodeless two-gap fit reproduces both the low-bias suppression and the coherence peaks, with $\Delta_i$ = 13.8 meV and $\Delta_o$ = 19.6 meV (Fig. 3c).

Fits containing a dominant nodal component fail to capture the flat bottom of the measured spectrum and leave excessive spectral weight near zero bias (Extended Data Fig. 4). The fitting comparison does not rule out weak anisotropy, matrix-element effects or orbital-selective lifetime broadening; rather, it shows that the principal local density of states in the rapidly transferred films is fully opened. This point is important because a V-shaped spectrum could otherwise be misread as a signature of nodal superconductivity. In the present samples, the U-shaped spectrum is reproducible in the oxygen-sufficient condition, whereas the V-shaped spectrum appears only after transfer conditions that are more prone to oxygen loss.

Magnetic-field-dependent spectra provide an additional check on the robustness of the local superconducting features. Under a perpendicular field of 2.2 T, the spectrum shows only weak changes and no obvious suppression of the gap magnitude (Extended Data Fig. 5). The weak field response is consistent with a field scale well below the large upper critical fields expected for compressively strained bilayer nickelate thin films; disorder and finite-thickness effects may further obscure local spectral changes[10,38]. These considerations do not alter the central observation that rapidly transferred films exhibit a reproducible, fully opened two-gap spectrum.

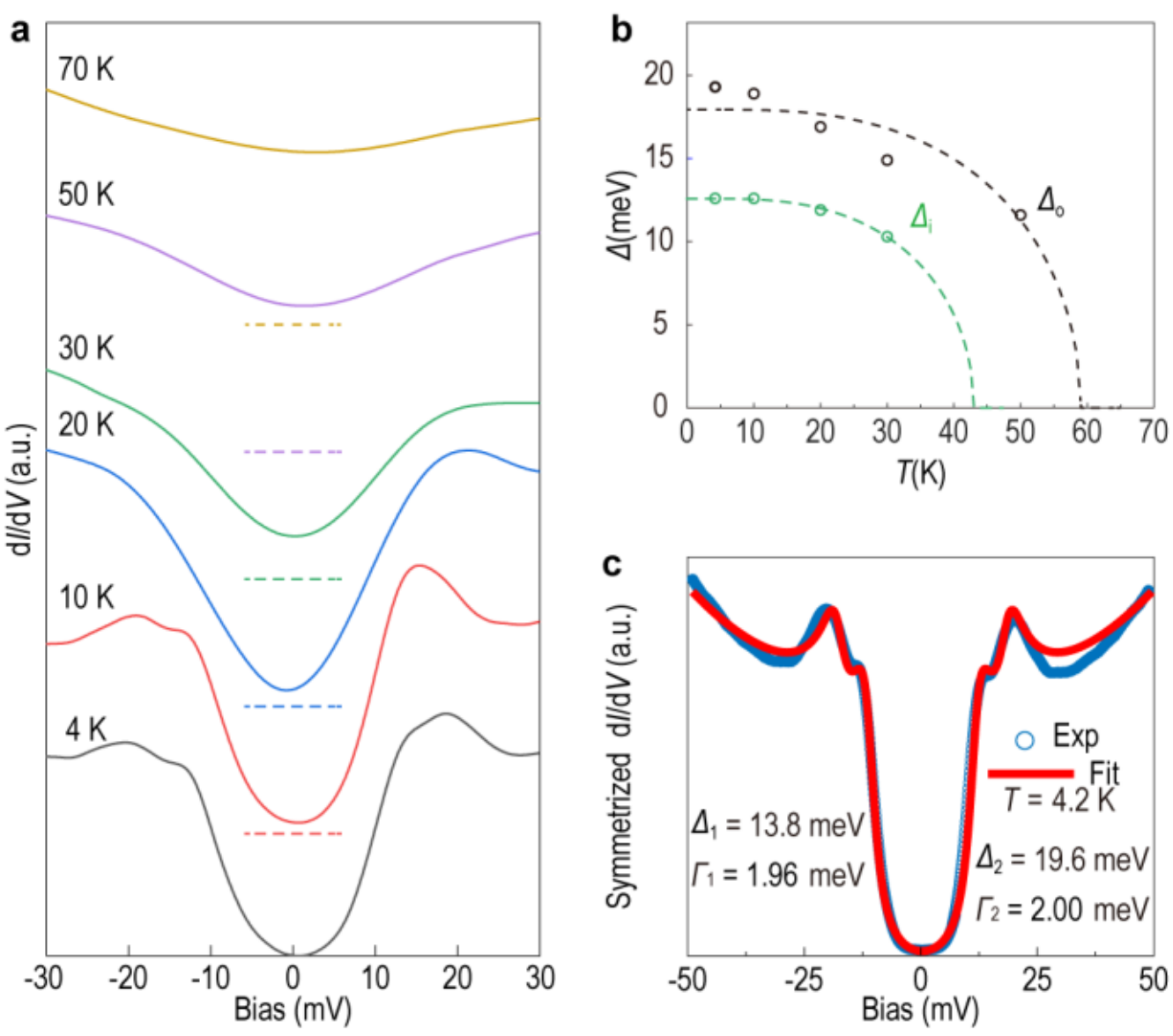

**Fig. 3 | Temperature evolution and two-gap modelling. a**, Tunnelling conductance spectra measured at different temperatures (set point: $V_s$ = 50 mV, $I_s$ = 300 pA, $V_{mod}$ = 1 mV). **b**, BCS-like fits (dashed curves) to $\Delta_i$ and $\Delta_o$ data extracted from peak positions in the corresponding $-d^3I/dV^3$ curves shown in Extended Data Fig. 3. **c**, Dynes-model fit to the normalized spectrum at 4.2 K. A nodeless two-gap fit with $\Delta_i$ = 13.8 meV and $\Delta_o$ = 19.6 meV reproduces the experimental spectrum.

## Comparison of U- and V-shaped tunnelling spectra

We found that the sample transfer time, defined as the time from removal from the low-temperature suitcase to arrival in the STM chamber, is positively correlated with oxygen loss and critically determines the tunnelling line shape. Transfer times shorter than 5 min consistently yielded U-shaped superconducting spectra (Fig. 2 and Fig. 4a), whereas transfer times longer than 10 min yielded only V-shaped gaps of about 16 meV (Fig. 4c and Extended Data Fig. 6). Over a wider bias range, the U-shaped spectra show an asymmetric suppression with edges at about +120 and -85 meV (Fig. 4b), while the V-shaped spectra exhibit a broad asymmetric dip with edges near +80 and -90 meV (Fig. 4d). This broad energy scale is comparable to density-wave energy scales reported in oxygen-sensitive $La_3Ni_2O_{7-\delta}$ systems[22–25,39–41]. Notably, samples with longer transfer times still displayed the $\sqrt{2} \times \sqrt{2}$ reconstruction and a resistive $T_c$ onset of 40–50 K. Thus, neither the reconstruction nor the transport measurement alone guarantees that the local tunnelling spectrum reflects the intrinsic superconducting gap.

The transfer-time dependence provides a built-in control experiment because the growth procedure and nominal structure remain the same while the oxygen-loss exposure changes. The coexistence of a preserved reconstruction with a V-shaped spectrum shows that the surface lattice order is more robust than the superconducting gap. The wide-energy dip in the V-shaped spectra therefore cannot be treated as a simple background; it is part of the oxygen-history-dependent electronic response. In this sense, the U-shaped and V-shaped spectra represent different surface conditions of the same material platform: an oxygen-sufficient condition in which a nodeless multigap superconducting spectrum is resolved, and a more oxygen-deficient condition in which density-wave-related spectral weight and degraded superconductivity are mixed.

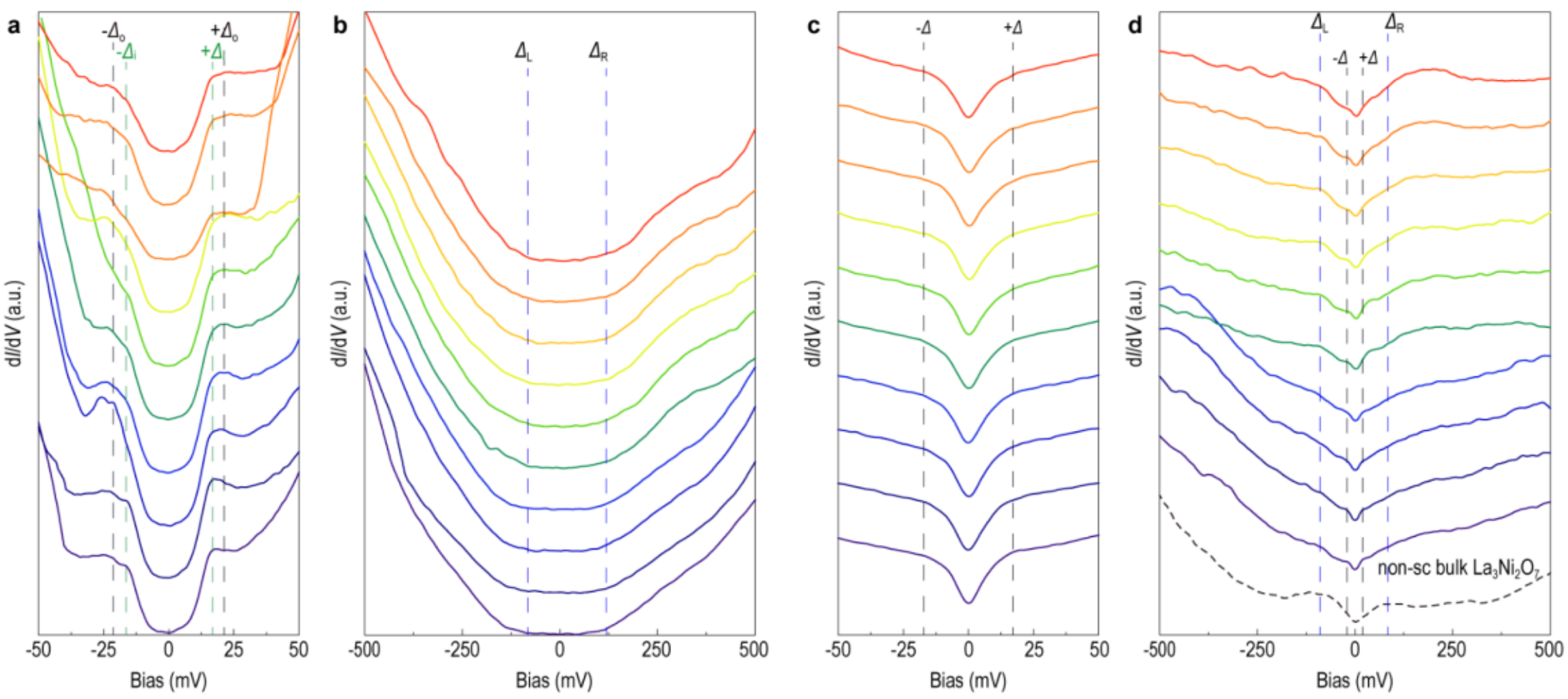

**Fig. 4 | Transfer-time dependence of U- and V-shaped tunnelling spectra. a**, Tunnelling conductance spectra (set point: $V_s$ = 50 mV, $I_s$ = 300 pA, $V_{mod}$ = 1 mV) exhibiting a U-shaped superconducting gap, with outer and inner gap energies of about 20 and 12–15 meV, respectively. **b**, Corresponding wide-energy-range spectra (set point: $V_s$ = 500 mV, $I_s$ = 300 pA, $V_{mod}$ = 5 mV), showing a broad asymmetric suppression with spectral-weight loss between approximately -85 and +120 meV. **c**, Tunnelling conductance spectra from a

sample with a longer transfer time, showing a V-shaped gap of about 16 meV. **d**, Corresponding wide-energy-range spectra, showing a broad dip with edges near -90 and +80 meV. The bottom black dashed curve represents the non-superconducting bulk $La_3Ni_2O_7$ spectrum, adapted from ref.[41].

## Discussion

Our STM/STS measurements establish that the local gap line shape in bilayer nickelate films is controlled by the surface oxygen content. With rapid cryogenic UHV transfer, the reconstructed surface exhibits a homogeneous U-shaped double gap with strongly suppressed zero-bias conductance. This fully opened spectrum is well described by nodeless two-gap Dynes fits and is inconsistent with a dominant nodal gap. By contrast, longer transfer through the room-temperature stages produces V-shaped spectra, even though the $\sqrt{2} \times \sqrt{2}$ reconstruction and a 40–50 K transport transition remain observable. The central implication is that the line shape of the local superconducting gap is more sensitive to oxygen loss than either the surface reconstruction or the macroscopic transport transition.

This comparison places an important constraint on the interpretation of V-shaped spectra. A V-shaped tunnelling gap should not be taken, by itself, as evidence for d-wave superconductivity. Oxygen vacancies are known to reorganize ligand holes in $La_3Ni_2O_{7-\delta}$[22], and oxygen-deficient films show spin- and charge-related superlattice responses[23]. Moreover, density-wave energy scales in $La_3Ni_2O_{7-\delta}$ are comparable to the broad dip observed in our wide-energy-range spectra[39–41]. The non-superconducting bulk $La_3Ni_2O_7$ spectrum shows a gaplike suppression with edges close to this energy range[42]. In our films, however, the corresponding broad dip emerges in the oxygen-loss-prone transfer condition. We therefore treat it as an oxygen-loss- and density-wave-sensitive background mixed with a degraded superconducting gap. The observed gap scale near 20 meV appears robust across local and point-contact probes, whereas the lower-energy line shape and residual low-bias conductance are much more sensitive to oxygen loss and surface processing[43]. This comparison supports a nodeless superconducting gap in the oxygen-sufficient condition, while also explaining why V-shaped spectra can appear in nominally superconducting nickelate samples.

The oxygen-stoichiometry dependence also helps reconcile local spectroscopy with the broader phase diagram. Interstitial oxygen order competes with superconductivity in $La_2PrNi_2O_{7+\delta}$[44], whereas oxygen deficiency can promote density-wave-related electronic reconstruction[22–25,39–41]. Superconductivity therefore appears in a window in which ligand-hole density, interlayer coupling and magnetic correlations are balanced. In this window, the U-shaped spectra reveal a nodeless multigap superconducting state. Outside it, V-shaped spectra can survive together with structural reconstruction and a transport transition. This picture also explains why the transfer protocol must be reported alongside the growth recipe for STM studies of nickelate films.

In summary, atomic-resolution STM/STS on oxygen-sufficient 1.5-unit-cell $(La,Pr)_3Ni_2O_7$

films reveals a homogeneous U-shaped double gap. Oxygen loss produces degraded V-shaped gap spectra with a broad dip on density-wave energy scales, even when the surface reconstruction and a 40–50 K transport superconducting transition remain. Together, these results provide an atomic-scale visualization of the intrinsic superconducting gap in bilayer nickelate thin films and establish oxygen-controlled transfer as a prerequisite for resolving pairing symmetry in Ruddlesden–Popper nickelates.

## Methods

Growth of $(La,Pr)_3Ni_2O_7$ films

All samples were grown on as-received $SrLaAlO_4$ (001) substrates using the gigantic-oxidative atomic-layer-by-layer epitaxy method[35]. During epitaxial growth of the $La_{1.35}Pr_{1.65}Ni_2O_7$ thin films, the $(La,Pr)O_x$ and $NiO_y$ targets were alternately ablated. The stoichiometry was achieved by calibrating the pulse counts for each target to ensure full coverage of the corresponding atomic layer. Further details of the pulse-count calibration and growth optimization can be found in ref.[45]

The 1.5 UC $La_{1.35}Pr_{1.65}Ni_2O_7$ film was synthesized by following the deposition sequence $(La,Pr)O$-$NiO_2$-$(La,Pr)O$-$NiO_2$-$(La,Pr)O$. The termination was obtained by depositing an additional sequence of $(La,Pr)O$-$NiO_2$-$(La,Pr)O$-$NiO_2$ on top of the 1 UC film.

The targets were rotated at a constant linear velocity of 5 mm/s to ensure uniform target erosion and maintain the compositional consistency of the deposited material. A buffer layer with the sequence $(La,Pr)O$-$NiO_2$-$(La,Pr)O$ was epitaxially grown before the $La_{1.35}Pr_{1.65}Ni_2O_7$ film to alleviate the interfacial discontinuity between the substrate and the film[45–46].

The $La_{0.45}Pr_{0.55}O_x$ and $NiO_y$ targets were synthesized from a stoichiometric mixture of $La_2O_3$ and $Pr_6O_{11}$ powders and pure NiO powder, respectively. Films were grown at about 850 °C with a laser fluence of 1.4–1.7 J $cm^{-2}$, and the pulsed-laser repetition rate was 4 Hz. The deposition was performed in a mixed atmosphere of $O_3$ and $O_2$ with a total pressure of 10 Pa, where the $O_3$ concentration was 10%. After deposition, the samples were rapidly cooled and transferred into the low-temperature UHV suitcase once the sample temperature was below 150 K.

STM/STS measurements

STM/STS experiments were performed in a commercial USM-1200 ultrahigh-vacuum MBE-STM combined system at 4.2 K. Immediately after PLD/GAE growth, the films were loaded into a low-temperature UHV suitcase held below 150 K with a base pressure below $5 \times 10^{-9}$ mbar. The suitcase was then docked to the STM load-lock/preparation chamber, and the sample passed sequentially through the low-temperature suitcase, the preparation chamber and the low-temperature STM chamber. We define the transfer time, used here as an operational measure of oxygen loss, as the elapsed time from taking the sample out of the suitcase environment at the STM system to loading it into the STM chamber. Samples with transfer times shorter than 5 min showed U-shaped superconducting gaps. Samples with transfer times longer than 10 min showed only V-shaped gaps, although such samples could still show the $\sqrt{2} \times \sqrt{2}$ reconstruction and a 40–50 K transport transition.

STM tips were prepared by electrochemical etching of tungsten wire and calibrated on Au(111) before measurements. Differential conductance spectra were acquired using a standard lock-in technique.

STM topographic image processing

To enhance the visibility of lattice-resolved structural features in the STM topographic image shown in the lower panel of Fig. 2a, we performed Fourier-filtering analysis on the raw data. Specifically, the raw STM image was first transformed into reciprocal space via fast Fourier transform (FFT). The central region of the FFT pattern, which corresponds to topographic undulations and defects, was then suppressed. Finally, an inverse FFT was applied to the remaining signal to reconstruct a spatially resolved image that highlights the lattice-related contributions.

Fitting Superconducting Gaps Using the Dynes Function

To quantitatively analyze the superconducting gap structure, the tunnelling spectra were fitted using the Dynes model, which describes the quasiparticle density of states with finite lifetime broadening. Assuming a metallic tip with a constant density of states near the Fermi level, the differential conductance measured by STS can be expressed as:

$$G(V) \propto \frac{\mathrm{d}}{\mathrm{dv}} \left( \int_{-\infty}^{+\infty} d\varepsilon \int_{0}^{2\pi} d\theta \; [f(\varepsilon) - f(\varepsilon + eV)] \cdot \mathrm{Re}\left( \frac{\varepsilon + eV + i\Gamma}{\sqrt{(\varepsilon + eV + i\Gamma)^2 - \Delta^2(\theta)}} \right) \right)$$

where $f(\varepsilon, T)$ is the Fermi-Dirac distribution function, $\Delta(\theta)$ denotes the superconducting gap function, and Gamma is the Dynes broadening parameter associated with finite quasiparticle lifetime effects. The angular variable $\theta$ describes the momentum dependence of the superconducting gap along the Fermi surface. In the fitting procedure, the effective electronic temperature and Gamma were treated as fitting parameters. The calculated spectra were fitted directly to the experimental STS data to extract the superconducting gap magnitude and its anisotropy. The best fits to the data and the fitting parameters are shown in Fig. 3c and Extended Data Fig. 4.

## Data Availability

All data that support the findings of this study are available from the corresponding author on reasonable request.

## Acknowledgements

This work is supported by the National Key Research and Development Program of China (grant nos. 2024YFA1408102, 2024YFA1408101 and 2022YFA1403101), the National Natural Science Foundation of China (grant nos. 12404157, 92565303, 92265112, 12374455, 52388201 and 12404171), the Guangdong Major Project of Basic Research (2025B0303000004), the Quantum Science Strategic Initiative of Guangdong Province (grant nos. GDZX2501001, GDZX2401004 and GDZX2201001), the Guangdong Project (Grant No. 2024QN11X176), the Municipal Funding Co-Construction Program of Shenzhen (grant nos. SZZX2401001 and SZZX2301004) and the Science and Technology Program of Shenzhen (grant no. KQTD20240729102026004), Station of Quantum

Materials.

## Author contributions

Q.-K.X., J.-F.J. and Y.S. supervised the entire project. Y.C. performed thin film growth under the supervision of G.Z. and Z.C.; L.X. and X.W. were responsible for sample transfer; X.W. and C.D. conducted the STM measurements under the supervision of Y.S. and C.D.; C.D., J.-J.M., Z.C., Y.S. and X.W. wrote the manuscript with input from all authors.

## Competing interests

The authors declare no competing interests.